\documentclass[10pt,aps,prd,floatfix,notitlepage,showpacs]{revtex4-1}
\usepackage{amsmath,amssymb}
\usepackage{graphicx}

\begin{document}

\title{Modified teleparallel theories of gravity}

\author{Sebastian Bahamonde}
\email{sebastian.beltran.14@ucl.ac.uk}

\author{Christian G. B\"ohmer}
\email{c.boehmer@ucl.ac.uk}

\author{Matthew Wright}
\email{matthew.wright.13@ucl.ac.uk}

\affiliation{Department of Mathematics, University College London, Gower Street, London, WC1E 6BT, United Kingdom}

\date{\today}

\begin{abstract}
We investigate modified theories of gravity in the context of teleparallel geometries. It is well known that modified gravity models based on the torsion scalar are not invariant under local Lorentz transformations while modifications based on the Ricci scalar are. This motivates the study of a model depending on the torsion scalar and the divergence of the torsion vector. We derive the teleparallel equivalent of $f(R)$ gravity as a particular subset of these models and also show that this is the unique theory in this class that is invariant under local Lorentz transformation. Furthermore one can show that $f(T)$ gravity is the unique theory admitting second order field equations.
\end{abstract}

\maketitle

\section{Introduction}

General Relativity is a very successful theory in excellent agreement with observations. However, the theory faces some challenges which are often summarised as the dark matter and the dark energy problem. The dark matter problem manifests itself, for instance, in flattened galactic rotation curves. Dark matter is an important ingredient for the dynamics of the entire universe and accounts for approximately 27\% its matter content, with dark energy making up about 68\%, with the remainder being ordinary matter. On the other hand, dark energy is responsible for the observed accelerated expansion of the universe. In principle one could accept the cosmological constant $\Lambda$ as an additional ingredient of physics, however, this causes substantial problems once the cosmological term is interpreted as a vacuum expectation value~\cite{Martin:2012bt}. 

Modifications of General Relativity (GR) started being considered almost immediately after the formulation of the theory. Many of those early studies were concerned with incorporating electromagnetism into the new geometrical framework, with advances in other branches of theoretical physics motivating a large variety of models. One such approach is based on a geometrical result going back to Weitzenb\"ock who observed that it is always possible to define a specific connection such that the space is globally flat. The geometrical framework is a manifold with curvature and torsion equipped with the so-called  Weitzenb\"ock connection. This forms the basis of what is now called the teleparallel equivalent of general relativity (TEGR), see e.g.~\cite{Obukhov:2002tm,Arcos:2005ec,Maluf:2013gaa,Weit,Cho,Cho2,Hayashi,Hay,Kop,Hammond:2002rm,Obukhov1} and also~\cite{Aldrovandi:2013wha}. In the standard GR framework, the metric contains the gravitational potentials which are responsible for the curvature of spacetime. On the other hand, in the teleparallel framework the gravitational fields are represented by the torsion tensor with the curvature not being important. While both formulations are equivalent, their interpretations are quite different. For instance, both formulations are invariant under local Lorentz transformations, however, in GR all geometrical quantities are naturally Lorentz scalars while in TEGR expressions typically depend on the chosen frame. 

In this paper we are interested in the classes of models known as $f(R)$ gravity and $f(T)$ gravity~\cite{Sotiriou:2008rp,DeFelice:2010aj,Capozziello:2011et,Nojiri:2010wj,Ferraro:2006jd,Bengochea:2008gz,Li:2010cg,Sotiriou:2010mv}. It is well known that $f(T)$ gravity is not invariant under local Lorentz transformations because the torsion scalar $T$ is not an invariant under them either, the Ricci scalar and the torsion scalar differ by a total derivative term. However, the resulting $f(T)$ gravity theory is a second order theory, unlike $f(R)$ gravity which contains fourth derivatives. By taking a fresh look at these models we derive the teleparallel equivalent of $f(R)$ gravity as a particular subset of models depending on the torsion scalar and a boundary term. We establish that this is the unique theory in this class that is invariant under local Lorentz transformation. Furthermore we can show that $f(T)$ gravity is the unique theory admitting second order field equations.

Our paper is organised as follows: Section~\ref{sec:tegr} gives a brief introduction to teleparallel gravity. Section~\ref{sec:ftbgravity} defines our theory and discusses its main features. We conclude in Section~\ref{sec:con}. We work with metric with signature $(-,+,+,+)$, Latin indices indicate tangents space coordinates whereas Greek indices correspond to spacetime coordinates.

\section{Teleparallel gravity}
\label{sec:tegr}

Let us begin by briefly introducing teleparallel gravity and its generalisation to $f(T)$ gravity. Our dynamical variables are the tetrad fields $e^{a}_\mu$, and inverse tetrads $E^{\mu}_{a}$ where Latin indices indicate tangents space coordinates whereas Greek indices correspond to spacetime coordinates.

The fundamental relationships between the metric $g_{\mu\nu}$, the inverse metric $g^{\mu\nu}$ and the tetrads and inverse tetrads are
\begin{align}
  g_{\mu\nu} &= e^{a}_{\mu} e^{b}_{\nu} \eta_{ab} \,, \\
  g^{\mu\nu} &= E^{\mu}_{a} E^{\nu}_{b} \eta^{ab} \,,
\end{align} 
where $\eta_{ab}$ is the Minkowski metric with signature $(-,+,+,+)$. The tetrad $e^{a}_\mu$ and the inverse tetrad $E^{\mu}_{a}$ satisfy
\begin{align}
  E_{m}^{\mu} e_{\mu}^{n} &= \delta^{n}_{m} \,,
  \label{deltanm} \\
  E_{m}^{\nu} e_{\mu}^{m} &= \delta^{\nu}_{\mu} \,.
  \label{deltamunu}
\end{align}
We define $e$ to be the determinant of the tetrad $e^a_\mu$, which is equivalent to the volume element of the metric, so that $e=\sqrt{-g}$ where $g$ is the determinant of the metric. In what follows, the conventions of~\cite{Maluf:2013gaa} are used.

General Relativity is a metric theory of gravity without additional geometrical objects being considered. The Riemann curvature tensor is constructed from the Levi-Civita connection which then gives rise to the standard Einstein tensor and the well known Einstein field equations. However, there exists an equivalent formulation based on a globally flat space where gravity is described by torsion instead of curvature. That this is indeed possible is not trivial and is based on the work of Weitzenb\"ock who noted that by choosing the connection in a specific way it is possible to ensure that space is indeed globally flat.

In the teleparallel formulation of General Relativity one works with the so-called Weitzenb\"ock connection. To begin with we define the object $W_{\mu}{}^{a}{}_{\nu}$ by
\begin{align}
  W_{\mu}{}^{a}{}_{\nu} = \partial_{\mu}e^{a}{}_{\nu} \,.
\end{align}
The torsion tensor is the antisymmetric part of $W_{\mu}{}^{a}{}_{\nu}$ so that
\begin{align}
  T^{a}{}_{\mu\nu} &= W_{\mu}{}^{a}{}_{\nu} - W_{\nu}{}^{a}{}_{\mu} =
  \partial_{\mu} e_{\nu}^{a} - \partial_{\nu}e_{\mu}^{a} \,,
\end{align}
or in terms of spacetime indices as
\begin{align}
  T^{\lambda}{}_{\mu\nu} = E_{a}^{\lambda} T^{a}{}_{\mu\nu} \,.
\end{align}

One can relate the Levi-Civita connection ${}^0 \Gamma$ and the Weitzenb\"ock connection as follows
\begin{align}
  W_{\lambda}{}^{\mu}{}_{\rho} = {}^0 \Gamma^{\mu}_{\lambda\rho} + K_{\lambda}{}^{\mu}{}_{\rho} \,,
\end{align}
where $K$ is the contortion tensor which in turn can be expressed using the torsion tensor as
\begin{align}
	2K_{\mu}\,^{\lambda}\,_{\nu}&=T^{\lambda}\,_{\mu\nu}-T_{\nu\mu}\,^{\lambda}+T_{\mu}\,^{\lambda}\,_{\nu}.
\end{align}
It is clear that the contortion tensor $K_{\lambda}{}^{\mu}{}_{\rho}$ is antisymmetric in its last two indices. One also defines the torsion vector $T_{\mu}$ as the following contraction
\begin{align}
  T_{\mu}=T^{\lambda}{}_{\lambda\mu} \,.
\end{align}
 
Let us calculate the Ricci scalar of the Levi-Civita connection in terms of torsion. One arrives at the following relation
\begin{align}
  e R(e) = -e\left(\frac{1}{4}T^{abc}T_{abc}+\frac{1}{2}T^{abc}T_{bac}-T^aT_a\right) + 
  2\partial_\mu (e T^\mu) \,,
  \label{ricci1}
\end{align}
see also~\cite{Maluf:2013gaa}. This can be divided by $e$ and yields
\begin{align}
  R(e) = -\left(\frac{1}{4}T^{abc}T_{abc}+\frac{1}{2}T^{abc}T_{bac}-T^aT_a\right) + 
  \frac{2}{e}\partial_\mu (e T^\mu) \,,
\end{align}
which can also be written as 
\begin{align}
  R(e) = - S^{abc}T_{abc} + \frac{2}{e}\partial_\mu (e T^\mu) \,,
  \label{ricciS}
\end{align}
where the tensor $S$ is defined as follows
\begin{align}
  S^{abc} = \frac{1}{4}(T^{abc}-T^{bac}-T^{cab})+\frac{1}{2}(\eta^{ac}T^b-\eta^{ab}T^c) \,.
\end{align}
Its form in spacetime coordinates is given by
\begin{align}
  2S_{\sigma}{}^{\mu\nu} = K_{\sigma}{}^{\mu\nu} - \delta^{\mu}_{\sigma}T^{\nu} + 
  \delta^{\nu}_{\sigma}T^{\mu} \,.
  \label{S}
\end{align}

Frequently the combination $S^{abc}T_{abc}$ is referred to simply as the torsion scalar $T$. This results in a neat form of (\ref{ricciS}) which then reads
\begin{align}
  R(e) = - T + \frac{2}{e}\partial_\mu (e T^\mu) \,,
  \label{ricciT}
\end{align}
and forms the principal starting point for teleparallel gravity. By definition, the Ricci scalar is invariant under local Lorentz transformations. This cannot be said for the torsion scalar $T$ or the boundary term, while the particular combination $T-B$ is invariant, the individual terms are not. As we will study the boundary term in some detail we introduce the notation
\begin{align}
  B = \frac{2}{e}\partial_\mu (e T^\mu) = 2 \nabla_\mu T^\mu \,.
  \label{B}
\end{align}

The action of the teleparallel equivalent of general relativity (TEGR) is given by
\begin{align}
  S_{\rm TEGR} = \int T e\, d^4x \,,
\end{align}
and the usual Einstein-Hilbert action of general relativity is given by
\begin{align}
  S_{\rm GR} = \int R \sqrt{-g} \, d^4x \,.
\end{align}
Identity~(\ref{ricciT}) shows that the actions $S_{\rm TEGR}$ and $S_{\rm GR}$ only differ by a total derivative term which implies that the field equations derived from either of the two actions are equivalent. Clearly, both theories are also invariant under local Lorentz transformations.

A well studied modification of GR is to consider $f(R)$ gravity~\cite{Sotiriou:2008rp,DeFelice:2010aj,Capozziello:2011et} where $f$ is an arbitrary (sufficiently smooth) function of the Ricci scalar
\begin{align}
  S_{f(R)} = \int f(R) \sqrt{-g} \, d^4x \,.
\end{align}
Recall that the Ricci scalar depends on second derivatives of the metric tensor. Hence variations with respect to the metric will require integration by parts twice which will result in terms of the form $\nabla_{\mu} \nabla_{\nu} F$ where $F=f'(R)$, making the theory fourth order. 

In analogy one can consider $f(T)$ gravity 
\begin{align}
  S_{f(T)} = \int f(T) e \, d^4x \,.
\end{align}
in the TEGR framework~\cite{Bengochea:2008gz}. Since the torsion scalar $T$ only depends on the first derivatives of the tetrads, this theory is a second order theory. However, since $f(T)$ does not differ from $f(R)$ by a total derivative term, these theories are no longer equivalent. Moreover, since $T$ itself is not invariant under local Lorentz transformations, $f(T)$ gravity is also not locally Lorentz invariant~\cite{Li:2010cg,Maluf:2013gaa}. Hence, there is a trade-off between second order field equations and local Lorentz invariance.

\section{$f(T,B)$ gravity}
\label{sec:ftbgravity}

We will now consider a general framework which includes both $f(R)$ gravity and $f(T)$ gravity as special sub-cases. Inspired by the above discussion, we define the action
\begin{align}
  S_{\rm TB} = \int 
  \left[ 
    \frac{1}{\kappa}f(T,B) + L_{\rm m}
  \right] e\, d^4x \,,\label{action}
\end{align}
where $f$ is a function of both of its arguments and $L_{\rm m}$ is a matter Lagrangian. 

Variations of the action with respect to the tetrad gives
\begin{align}
  \delta S_{\rm TB} = \int 
  \left[ 
    \frac{1}{\kappa}
    \Big(
    f(T,B)\delta e + e f_{B}(T,B) \delta B + e f_{T}(T,B) \delta T
    \Big) 
    + \delta(e L_{\rm m})
  \right] 
  \, d^4x \,,
  \label{action2}
\end{align}
where
\begin{align}
  ef_{B}(T,B)\delta B & = \Big[2eE_{a}^{\nu}\nabla^{\lambda}\nabla_{\mu}f_{B}-2eE_{a}^{\lambda}\Box f_{B}-Bef_{B}E_{a}^{\lambda}-4e(\partial_{\mu}f_{B})S_{a}\,^{\mu\lambda}\Big]\delta e_{\lambda}^{a} \,,
  \label{deltaB}\\
  ef_{T}(T,B)\delta T & = \Big[-4e(\partial_{\mu}f_{T})S_{a}\,^{\mu\lambda}-4\partial_{\mu}(e S_{a}\,^{\mu\lambda})f_{T}+4ef_{T}T^{\sigma}\,_{\mu a}S_{\sigma}\,^{\lambda\mu}\Big]\delta e^{a}_{\lambda} \,,
  \label{deltaT}\\
  f(T,B)\delta e & = e f(T,B)E_{a}^{\lambda} \delta e^{a}_{\lambda} 
  \label{deltae1} \,.
\end{align}
The variations~(\ref{deltaT}),(\ref{deltae1}) are the standard variations that lead to $f(T)$ gravity. A detailed derivation of the $\delta B$ variation~(\ref{deltaB}) is presented in the Appendix. The energy momentum tensor is defined as follows
\begin{align}
  \Theta^{\lambda}_{a}=\frac{1}{e} \frac{\delta (eL_m)}{\delta e^{a}_{\lambda}} \,.
\end{align}
Putting everything together, we find that the field equations are given by
\begin{multline}
  2eE_{a}^{\lambda}\Box f_{B}-2eE_{a}^{\sigma}\nabla^{\lambda}\nabla_{\sigma}f_{B}+
  eBf_{B}E_{a}^{\lambda} + 4e\Big[(\partial_{\mu}f_{B})+(\partial_{\mu}f_{T})\Big]S_{a}{}^{\mu\lambda} 
  \\
  +4\partial_{\mu}(e S_{a}{}^{\mu\lambda})f_{T}-4ef_{T}T^{\sigma}{}_{\mu a}S_{\sigma}{}^{\lambda\mu}-
  e f E_{a}^{\lambda} = 16\pi e \Theta_{a}^{\lambda}.
\end{multline}
And contracting this with $e^{a}_{\nu}$ we arrive at the field equations in spacetime indices only
\begin{multline}
  2e\delta_{\nu}^{\lambda}\Box f_{B}-2e\nabla^{\lambda}\nabla_{\nu}f_{B}+
  e B f_{B}\delta_{\nu}^{\lambda} + 
  4e\Big[(\partial_{\mu}f_{B})+(\partial_{\mu}f_{T})\Big]S_{\nu}{}^{\mu\lambda}
  \\
  +4e^{a}_{\nu}\partial_{\mu}(e S_{a}{}^{\mu\lambda})f_{T} - 
  4 e f_{T}T^{\sigma}{}_{\mu \nu}S_{\sigma}{}^{\lambda\mu} - 
  e f \delta_{\nu}^{\lambda} = 16\pi e \Theta_{\nu}^{\lambda} \,.
  \label{fieldeq}
\end{multline}
where $\Theta_{\nu}^{\lambda}=e^{a}_{\nu}\Theta_{a}^{\lambda}$ is the standard energy momentum tensor. In the following we will consider the limiting cases which give $f(T)$ gravity and $f(R)$ gravity, respectively. To be more precise, we actually derive the teleparallel equivalent of $f(R)$ gravity and show its equivalence with $f(R)$ gravity. 

\subsection{$f(T)$ gravity}

Let us begin with examining our field equation~(\ref{fieldeq}) when choosing the function $f$ to be independent of the boundary term. In order to match the sign convention employed, we simply set
\begin{align}
  f(T,B)= f(T) \,,
\end{align}
so that $f_B=0$. Doing this, we find
\begin{align}
  4e\Big[f_{TT}(\partial_{\mu}T)\Big]S_{\nu}{}^{\mu\lambda}+
  4e^{a}_{\nu}\partial_{\mu}(e S_{a}{}^{\mu\lambda})f_{T}-
  4ef_{T}T^{\sigma}{}_{\mu \nu}S_{\sigma}{}^{\lambda\mu}-
  ef\delta_{\nu}^{\lambda} = 16\pi e \Theta_{\nu}^{\lambda} \,,
  \label{fT}
\end{align}
which, as expected, are the standard $f(T)$ field equations. Let us make an important remark about this limit. One verifies immediately that this is the unique form of the function $f$ which will give second order field equations. Recall that linear terms in the boundary term $B$ do not affect the field equations. Therefore, the generic field equations contain terms of the form $\partial_\mu \partial_\nu f_b$ which are always of fourth order and can vanish if and only if $f_b$ is a constant, so that $f$ is linear in the boundary term.

Therefore, for a nonlinear function $f$, $f(T)$ gravity is the only possible second order modified theory of gravity constructed out of $R$, $T$ and $B$. As mentioned before, the price to pay is the violation of local Lorentz invariance. 

\subsection{$f(R)$ gravity}

Here we will show carefully how we recover standard $f(R)$ gravity in this model, and also find the teleparallel equivalent of $f(R)$ gravity. Starting with~(\ref{ricciT}), we have
\begin{align}
  R = -T + B \,,
\end{align}
which suggests to consider our function $f$ to be of the particular form
\begin{align}
  f(T,B) = f(-T+B) \,.
\end{align}
We also introduce the standard notation for the derivative of $f$ from $f(R)$ gravity
\begin{align}
  F(R) = f'(-T+B) = -f_{T} = f_{B} \,.
\end{align}

Inserting this form of function into our general $f(T,B)$ field equation~(\ref{fieldeq}) leads to  the following field equations
\begin{align}
  2e\delta_{\nu}^{\lambda}\Box F-2e\nabla^{\lambda}\nabla_{\nu}F+eBF\delta_{\nu}^{\lambda}
  -4e^{a}_{\nu}\partial_{\mu}(e S_{a}{}^{\mu\lambda})F+4eFT^{\sigma}{}_{\mu \nu}S_{\sigma}{}^{\lambda\mu}-
  e f \delta_{\nu}^{\lambda} = 16\pi e \Theta_{\nu}^{\lambda} \,.
  \label{fRtele}
\end{align}
This equation gives us the teleparallel equivalent of $f(R)$ gravity (although for simplicity we have expressed this equation using covariant derivatives $\nabla$ of the Levi-Civita connection, these can easily be rewritten in the teleparallel framework using the relation $\nabla_\mu V^{\mu}=\frac{1}{e}\partial_\mu (e V^\mu)$). As this is not an obvious observation, let us prove this statement by rewriting the field equations in their usual way based on the Ricci tensor and metric tensor.

We can rewrite the fourth term in~(\ref{fRtele}) as
\begin{align}
  4e^{a}_{\nu}\partial_{\mu}(eS_{a}{}^{\mu\lambda}) &= 2\partial_{\mu}(eK_{\nu}{}^{\mu\lambda})-2\partial_{\nu}(eT^{\lambda})+eB\delta^{\lambda}_{\nu} 
  + 4eS_{\sigma}{}^{\lambda\mu}W_{\mu}{}^{\sigma}{}_{\nu} \,.
\end{align}
Inserting this back into~(\ref{fRtele}) gives
\begin{eqnarray}
  2e\delta_{\nu}^{\lambda}\Box F-2e\nabla^{\lambda}\nabla_{\nu}F-
  2F\partial_{\mu}(eK_{\nu}\,^{\mu\lambda})+2F\partial_{\nu}(eT^{\lambda})-
  4 e F S_{\sigma}{}^{\lambda\mu}W_{\nu}{}^{\sigma}{}_{\mu} -
  e f \delta_{\nu}^{\lambda} = 16\pi e \Theta_{\nu}^{\lambda} \,.
  \label{fRtele2}
\end{eqnarray}
Next, we need to replace the torsion components with curvature. The Ricci tensor satisfies the identity
\begin{align}
  ^{0}R_{\mu\nu} = \nabla_{\nu}K_{\lambda}{}^{\lambda}{}_{\mu} - 
  \nabla_{\lambda}K_{\nu}{}^{\lambda}{}_{\mu} +
  K_{\lambda}{}^{\rho}{}_{\mu}K_{\nu}{}^{\lambda}{}_{\rho} -
  K_{\lambda}{}^{\lambda}{}_{\rho}K_{\nu}{}^{\rho}{}_{\mu} \,.
\end{align}
We can rewrite this to derive the following identity
\begin{align} 
  ^{0}R^{\lambda}_{\nu}  = \frac{1}{e}\Big(
  \partial_{\sigma}(eK_{\nu}{}^{\lambda\sigma})+\partial_{\nu}(eT^{\lambda})
  \Big) - 2S_{\sigma}{}^{\lambda\mu}W_{\nu}{}^{\sigma}{}_{\mu} \,.
\label{eq10B}
\end{align}
Using this final identity~(\ref{eq10B}), it is then easy to see that the field equations reduce to the $f(R)$ field equations in standard form
\begin{align}
  F R_{\mu\nu} - \frac{1}{2} f g_{\mu\nu} + 
  g_{\mu\nu}\Box F - \nabla_{\mu}\nabla_{\nu}F = 8\pi \Theta_{\mu\nu} \,,
\end{align}
where $\Theta_{\mu\nu}$ is the energy-momentum tensor. Thus we conclude that equation~(\ref{fRtele}) is the teleparallel equivalent of $f(R)$ gravity. 

\subsection{Lorentz invariance}

As in the previous subsection, let us rewrite our general field equation in a covariant form in terms of the Einstein tensor and the metric. If we insert the expression for the Ricci tensor~(\ref{eq10B}) into the field equation~(\ref{fieldeq}) we find
\begin{multline}
  2e\delta_{\nu}^{\lambda}\Box f_{B}-2e\nabla^{\lambda}\nabla_{\nu}f_{B} + 
  e B f_{B}\delta_{\nu}^{\lambda} + 4 e 
  \Big[(f_{BB}+f_{BT})(\partial_{\mu}B)+(f_{TT}+f_{BT})(\partial_{\mu}T)\Big]S_{\nu}{}^{\mu\lambda}
  \\
  +4e^{a}_{\nu}\partial_{\mu}(e S_{a}{}^{\mu\lambda})f_{T} - 
  4 e f_{T} T^{\sigma}{}_{\mu \nu}S_{\sigma}{}^{\lambda\mu} -
  ef\delta_{\nu}^{\lambda} = 16\pi e \Theta_{\nu}^{\lambda} \,.
  \label{eq8}
\end{multline}
Using the relation $R=-T+B=-T+2\partial_{\mu}T^{\mu}$ and $R_{\nu}^{\lambda} = G_{\nu}^{\lambda} + \frac{1}{2}(B-T) \delta_{\nu}^{\lambda}$, after some algebra, we can write the field equation in the following form
\begin{multline}
  H_{\mu\nu} := -f_{T}G_{\mu\nu}+g_{\mu\nu}\Box f_{B}-\nabla_{\mu}\nabla_{\nu}f_{B} +
  \frac{1}{2}(Bf_{B}+Tf_{T}-f)g_{\mu\nu}
  \\
  +2\Big[(f_{BB}+f_{BT})(\nabla_{\lambda}B)+(f_{TT}+f_{BT})(\nabla_{\lambda}T)\Big]S_{\nu}{}^{\lambda}{}_{\mu}
  = 8\pi \Theta_{\mu\nu} \,.
  \label{eq9}
\end{multline}
It is readily seen that if one considers the $f(T)$ limit, then this equation coincides with the covariant form of the $f(T)$ field equations presented in~\cite{Li:2010cg}, and we note that this equation is manifestly covariant. However, it is not in general invariant under local Lorentz transformations.  A necessary condition for the equation to be Lorentz invariant is for the antisymmetric part of the equation to be identically zero, so the coefficient of $S_{\nu}{}^{\lambda}{}_{\mu}$ must vanish identically, see for example~\cite{Li:2010cg}. Requiring this gives two conditions
\begin{align}
  f_{BB} + f_{BT} = 0\,, \quad \text{and} \quad
  f_{TT} + f_{BT} = 0\,,
\end{align}
which can be satisfied if we choose
\begin{align}
  f_T + f_B = c_1 \,,
\end{align}
where $c_1$ is a constant of integration. Solving this gives us a general $f$ of the form
\begin{align}
  f(T,B) = \tilde{f}(-T+B) + c_1 B = \tilde{f}(R) + c_1 B \,.
\end{align}
Since $B$ is a total derivative term, the resulting field equations are unchanged by terms linear in $B$. Hence, we can set $c_1=0$ without loss of generality. We already showed that an $f$ of this form simply reduces to the $f(R)$ field equations, which are manifestly Lorentz invariant. Hence we can conclude that the above field equations are Lorentz invariant if and only if they are equivalent to $f(R)$ gravity. Therefore, the teleparallel equivalent of $f(R)$ gravity is the only possible Lorentz invariant theory of gravity constructed out of $R$, $T$ and $B$. Conversely to the above, the price to pay is the presence of higher order derivative terms. 

\subsection{Conservation equations}

Requiring the matter action to be invariant under both local Lorentz transformations and infinitesimal coordinate transformations gives the condition that $T_{\mu\nu}$ is symmetric and divergence free
\begin{align}
  \nabla^{\mu} \Theta_{\mu\nu}=0 \,.
\end{align}
as shown in~\cite{Li:2010cg}. Hence we require the left-hand side of our field equations to also have this property. Let us show that this is indeed the case and that there is no need for this to be imposed as an extra (independent) condition.

For compactness, let us define the vector
\begin{align}
  X_\lambda=\Big[(f_{BB}+f_{BT})(\nabla_{\lambda}B)+(f_{TT}+f_{BT})(\nabla_{\lambda}T)\Big].
\end{align} 
Taking the covariant derivative of $H^{\mu\nu}$, we find after some simplification
\begin{align}
  \nabla^\mu H_{\mu\nu}=-\left[R_{\mu\nu}-\frac{1}{2}Bg_{\mu\nu}+2\nabla^\rho S_{\nu\rho\mu}\right]X^\mu.
\end{align}
Now using
\begin{align}
  R_{\mu\nu}=-2\nabla^\rho S_{\nu\rho\mu}+\frac{1}{2}Bg_{\mu\nu}-2S^{\rho\sigma}{}_\mu K_{\nu\sigma\rho},
\end{align} 
this simplifies to
\begin{align}
  \nabla^\mu H_{\mu\nu}=2S^{\rho\sigma}{}_\mu K_{\nu\sigma\rho} X^\mu.
\end{align}
However, we know that the energy momentum tensor is symmetric, and hence
\begin{align}
  H_{[\mu\nu]}=-S_{[\nu\mu]}\,^{\lambda} X_{\lambda}=0.
\end{align}
This implies
\begin{align}
  \nabla^\mu H_{\mu\nu}=2H^{[\rho\sigma]} K_{\nu\rho\sigma}=0,
\end{align}
which follows from $K$ being antisymmetric in its last two indices. This means that on shell the left-hand side of the field equations are conserved.

\section{Conclusions}
\label{sec:con}

\begin{figure}[!ht]
\centering
\includegraphics[width=0.48\textwidth]{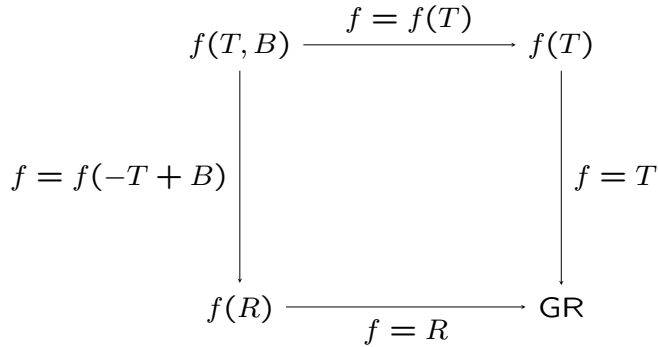}
\caption{Relationship between different modified gravity models and General Relativity.}
\label{fig1}
\end{figure}

The principal aim of this work was to complete our understanding of the relationship between different models of modified gravity in the context of $f(R)$ and $f(T)$ gravity. Our results can be visualised using Fig.~\ref{fig1}. The starting point is a gravitational action based on an arbitrary function $f(T,B)$ which depends on the torsion scalar and a torsion boundary term. If this function is assumed to be independent of the boundary term, one arrives at $f(T)$ gravity which we identified as the unique second order gravitational theory in this approach. Likewise, if the function takes the special form $f(-T+B)$, we find the teleparallel equivalent of $f(R)$ gravity. This theory is identified as the unique locally Lorentz invariant theory. Any other form of $f(T,B)$ will result in gravitational theories which are neither of second order nor locally Lorentz invariant. 

Based on these considerations, one could study some interesting physical models using the teleparallel framework. For instance, it would be interesting to couple a scalar field to this boundary term and study its cosmological applications. Clearly, one will be able to establish a direct link between Brans-Dicke type theories and their teleparallel counterparts~\cite{Geng:2011aj}. However, couplings to the boundary term will result in some new dynamics which will be interesting to study~\cite{Bahamonde:2015}.

As discussed in the above, neither the torsion scalar $T$ nor the boundary term $B$ are invariant under local Lorentz transformations. This might be useful when studying metric theories of gravity which already break this invariance, for instance theories containing a preferred direction, see for instance~\cite{Jacobson:2008aj}. One could reformulate these theories using the teleparallel framework and choose suitable coupling terms containing the boundary term so that the resulting theory becomes invariant under local Lorentz transformations. This will not be possible for any theory violating local Lorentz invariance, however, for some models this approach will yield a new invariant theory.

\begin{acknowledgments}
SB is supported by the Comisi{\'o}n Nacional de Investigaci{\'o}n Cient{\'{\i}}fica y Tecnol{\'o}gica (Becas Chile Grant No.~72150066).
\end{acknowledgments}

\appendix*

\section{Derivation of the field equations}

This appendix contains some details about the calculation which yields the variation~(\ref{deltaB}) of the Lagrangian, which corresponds to the variation of the additional dependence of $f$ on the boundary term $B$. This material is included to make this paper more self-contained and to make it easier to verify the results.

Performing this variation, it is found that 
\begin{align}
  ef_{B}(T,B)\delta B = -\Big(f_{B}B+2(\partial_{\mu}f_{B})T^{\mu}\Big)\delta e - 
  2e(\partial_{\mu}f_{B})\delta T^{\mu} \,,
  \label{transi1}
\end{align}
where we used that the torsion vector is given by 
\begin{align}
  T^{\mu} = g^{\mu\nu} T^{\sigma}{}_{\sigma\nu} =
  g^{\mu\nu} E_{a}^{\sigma} \Big(\partial_{\sigma}e_{\nu}^{a}-\partial_{\nu}e_{\sigma}^{a}\Big) \,.
\end{align}

Using that $\delta e=\frac{1}{2}g^{\mu\nu}\delta g_{\mu\nu}$ and $g^{\mu\nu}=\eta^{ab}E_{a}^{\mu}E_{b}^{\nu}$, it is easily shown that
\begin{align}
  \delta g^{\mu\nu} &= -\Big(g^{\nu\lambda}E_{a}^{\mu}+g^{\mu\lambda}E_{a}^{\nu}\Big)\delta e_{\lambda}^{a}
  \label{deltag} \,,\\
  \delta e &= eE_{a}^{\lambda}\delta e_{\lambda}^{a} \,.
  \label{deltae}
\end{align}

Now by varying~(\ref{deltamunu}) we find the relation of the variation of the inverse tetrad to the tetrad to be
\begin{align}
  \delta E_{m}^{\sigma} &= -E_{n}^{\sigma}E_{m}^{\mu}\delta e_{\mu}^{n} \,.
  \label{deltaE}
\end{align}
And by taking partial derivatives of~(\ref{deltamunu}), one can also find a similar relation for the partial derivatives of the inverse tetrad
\begin{align}
  \partial_{\nu} E_{m}^{\sigma}&=-E_{n}^{\sigma}E_{m}^{\mu}\partial_{\nu} e_{\mu}^{n} \,.
  \label{deltaDE}
\end{align}
Using~(\ref{deltag}) and~(\ref{deltaDE}), $\delta T^{\mu}$ can be written as
\begin{align}
  \delta T^{\mu} = -\Big(E_{a}^{\mu}T^{\lambda}+g^{\mu\lambda}T_{a}-T^{\lambda}{}_{a}{}^{\mu}\Big)\delta e^{a}_{\lambda}+
  g^{\mu\nu}E_{a}^{\lambda}\Big(\partial_{\lambda}\delta e^{a}_{\nu}-\partial_{\nu}\delta e^{a}_{\lambda}\Big) \,.
\end{align}
If we integrate by parts and disregard the boundary term, the last term on the right hand side of~(\ref{transi1}) becomes
\begin{align}
  e(\partial_{\mu}f_{B})\delta T^{\mu} &= 
  \Big[
    \partial_{\nu}\Big(E_{a}^{\lambda}(eg^{\mu\nu})(\partial_{\mu}f_{B})\Big)-
    \partial_{\nu}\Big(E_{a}^{\nu}(eg^{\mu\lambda})(\partial_{\mu}f_{B})\Big)
    \nonumber\\ &-
    e(\partial_{\mu}f_{B})
    \Big(E_{a}^{\mu}T^{\lambda}+g^{\mu\lambda}T_{a}+T^{\lambda}{}_{a}{}^{\mu}\Big)
  \Big]
  \delta e_{\lambda}^{a} \,.
  \label{transia1}
\end{align}
Using $\partial_{\lambda}e=e g^{\mu\nu}\partial_{\lambda}g_{\mu\nu}$ and the compatibility equation for the metric $\nabla_{\lambda}(g^{\mu\nu})=0$ we find
\begin{align}
  \partial_{\lambda}e &= eW_{\lambda}\,^{\rho}\,_{\rho} \,,
  \label{partiale}\\
  \partial_{\lambda}g^{\mu\nu} &= -\Big(W_{\lambda}{}^{\nu\mu}+W_{\lambda}{}^{\mu\nu}\Big) \,.
  \label{partialg}
\end{align}
The affine connection is
\begin{align}
  \Gamma^{\lambda}_{\mu\nu} &= W_{\mu}{}^{\lambda}{}_{\nu}-K_{\mu}{}^{\lambda}{}_{\nu} =
  W_{\nu}{}^{\lambda}{}_{\mu}-K_{\nu}{}^{\lambda}{}_{\mu} \,.
  \label{affine}
\end{align}
Using~(\ref{partiale}), (\ref{partialg}) and the equation above, the first term of~(\ref{transia1}) can be written in terms of covariant derivatives as
\begin{align}
	\partial_{\nu}\Big(E_{a}^{\lambda}(eg^{\mu\nu})(\partial_{\mu}f_{B})\Big) &= 
	eE_{a}^{\lambda}\Box f_{B}-e(\partial_{\mu}f_{B})
	\Big(E_{a}^{\lambda}W_{\nu}\,^{\mu\nu}-E_{a}^{\lambda}W^{\nu\mu}\,_{\nu}+W^{\mu\lambda}\,_{a}\Big) \,.
	\label{partialnu1}
\end{align}
Using the same idea, the second term of~(\ref{transia1}) becomes
\begin{align}
	\partial_{\nu}\Big(E_{a}^{\nu}(eg^{\mu\lambda})(\partial_{\mu}f_{B})\Big) &= 
	e E_{a}^{\nu}\nabla^{\lambda}\nabla_{\mu}f_{B}+e(\partial_{\mu}f_{B})
	\Big(g^{\mu\lambda}(W_{a}\,^{\nu}\,_{\nu}-W_{\nu}\,^{\nu}\,_{a})
	\nonumber\\ &-
	W_{a}\,^{\lambda\mu}-W_{a}\,^{\mu\lambda}+W^{\lambda\mu}\,_{a}-K^{\lambda\mu}\,_{a}\Big) \,.
	\label{partialnu2}
\end{align}

By inserting~(\ref{partialnu1}) and~(\ref{partialnu2}) into~(\ref{transia1}) we find
\begin{align}
  e(\partial_{\mu}f_{B})\delta T^{\mu} &= -
  \Big[
    e(\partial_{\mu}f_{B})
    \Big(E_{a}^{\mu}T^{\lambda}+g^{\mu\lambda}T_{a}+T^{\lambda}{}_{a}{}^{\mu}+
    g^{\mu\lambda}(W_{a}{}^{\nu}{}_{\nu}-W_{\nu}{}^{\nu}{}_{a})-W_{a}{}^{\lambda\mu}
    \nonumber\\ &-
    W_{a}{}^{\mu\lambda}+W^{\lambda\mu}{}_{a}-K^{\lambda\mu}{}_{a}-W_{a}{}^{\mu\lambda}+
    W^{\mu\lambda}{}_{a}+W_{\nu}{}^{\mu\nu}-W^{\nu\mu}{}_{\nu}\Big) -
    eE_{a}^{\lambda}\Box f_{B}
    \nonumber\\ &+ 
    e E_{a}^{\nu}\nabla^{\lambda}\nabla_{\mu}f_{B}
  \Big] \delta e_{\lambda}^{a} \,.
  \label{transia1a}
\end{align}
If we use the symmetry of the affine connection, i.e. Eq.~(\ref{affine}), we can simplify the equation as 
\begin{align}
  e(\partial_{\mu}f_{B})\delta T^{\mu} &= -
  \Big[
    e(\partial_{\mu}f_{B})
    \Big(E_{a}^{\mu}T^{\lambda}+W^{\lambda\mu}{}_{a}-W_{a}{}^{\mu\lambda}-K^{\lambda\mu}{}_{a}\Big)
    \nonumber\\ &-
    eE_{a}^{\lambda}\Box f_{B}+eE_{a}^{\nu}\nabla^{\lambda}\nabla_{\mu}f_{B}
  \Big] \delta e_{\lambda}^{a} \,.
  \label{transia1ab}
\end{align}
Now, by inserting this expression into~(\ref{transi1}) and using~(\ref{deltae}) we find
\begin{align}
  ef_{B}(T,B)\delta B &= 
  \Big[
    2eE_{a}^{\nu}\nabla^{\lambda}\nabla_{\mu}f_{B}-2eE_{a}^{\lambda}\Box f_{B}-
    Bef_{B}E_{a}^{\lambda}+2e(\partial_{\mu}f_{B})
    \Big(E_{a}^{\mu}T^{\lambda}-E_{a}^{\lambda}T^{\mu}
    \nonumber\\ &+
    W^{\lambda\mu}{}_{a}-W_{a}{}^{\mu\lambda}-K^{\lambda\mu}{}_{a}\Big)
  \Big] \delta e_{\lambda}^{a} \,.
\end{align}
Here we will introduce $2S_{a}{}^{\lambda\mu}=K_{a}{}^{\lambda\mu}+E_{a}^{\mu}T^{\lambda}-E_{a}^{\lambda}T^{\mu}$ to obtain
\begin{align}
  ef_{B}(T,B)\delta B &= 
  \Big[
    2eE_{a}^{\nu}\nabla^{\lambda}\nabla_{\mu}f_{B}-2eE_{a}^{\lambda}\Box f_{B}-
    Bef_{B}E_{a}^{\lambda}+2e(\partial_{\mu}f_{B})
    \Big(2S_{a}{}^{\lambda\mu}
    \nonumber\\ &-
    K_{a}{}^{\lambda\mu}+W^{\lambda\mu}{}_{a}-W_{a}{}^{\mu\lambda}-K^{\lambda\mu}{}_{a}\Big)
  \Big] \delta e_{\lambda}^{a} \,.
\end{align}
The last four terms on the right hand side are identically zero due to~(\ref{affine}). Thus, we obtain the final result which is
\begin{align}
  ef_{B}(T,B)\delta B &= 
  \Big[
    2eE_{a}^{\nu}\nabla^{\lambda}\nabla_{\mu}f_{B}-2eE_{a}^{\lambda}\Box f_{B}-
    Bef_{B}E_{a}^{\lambda}-4e(\partial_{\mu}f_{B})S_{a}{}^{\mu\lambda}
  \Big] \delta e_{\lambda}^{a} \,.
\end{align}

\end{document}